\documentclass{aa}
\usepackage{psfig}
\begin{document}
%\hyphenation{}               
%\renewcommand{\textfraction}{0.05}
%
%  Abbreviations
%
\def\etal {et al.}
\def\ie {i.\,e.}
\def\etseq {{\em et seq.}}
\def\vs {{it vs.}}
\def\perse {{it per se}}
\def\adhoc {{\em ad hoc}}
\def\eg {e.\,g.}
\def\etc {etc.}
\def\ccpers {\hbox{${\rm cm}^3{\rm s}^{-1}$}}
\def\vlsr {\hbox{${v_{\rm LSR}}$}}
\def\vhel {\hbox{${v_{\rm HEL}}$}}
\def\delv {\hbox{$\Delta v_{1/2}$}}
\def\TL {$T_{\rm L}$}
\def\TC {$T_{\rm c}$}
\def\TEX {$T_{\rm ex}$}
\def\TMB {$T_{\rm mb}$}
\def\TKIN {$T_{\rm kin}$}
\def\TREC {$T_{\rm rec}$}
\def\TSYS {$T_{\rm sys}$}
\def\TVIB {$T_{\rm vib}$}
\def\TROT {$T_{\rm rot}$}
\def\TDUST {$T_{\rm d}$}
\def\TASTAR {$T_{\rm A}^{*}$}
\def\TVIBST {$T_{\rm vib}^*$} 
\def\H0 {$H_{\rm o}$}
\def\mic {$\mu\hbox{m}$}
\def\micro {\mu\hbox{m}}
\def\SDOZ {\hbox{$S_{12\mu \rm m}$}}
\def\STWE {\hbox{$S_{25\mu \rm m}$}}
\def\SSIX {\hbox{$S_{60\mu \rm m}$}}
\def\SHUN {\hbox{$S_{100\mu \rm m}$}}
\def\solmass {\hbox{M$_{\odot}$}}
\def\solum {\hbox{L$_{\odot}$}}
\def\irlum {\hbox{$L_{\rm FIR}$}}
\def\ohlum {\hbox{$L_{\rm OH}$}}
\def\blum {\hbox{$L_{\rm B}$}}
\def\numd {\hbox{$n({\rm H}_2$)}}                   
\def\rhounit {$\hbox{M}_\odot\,\hbox{pc}^{-3}$}
\def\kms {\hbox{${\rm km\,s}^{-1}$}}
\def\kmsyr {\hbox{${\rm km\,s}^{-1}\,{\rm yr}^{-1}$}}
\def\kmsmpc {\hbox{${\rm km\,s}^{-1}\,{\rm Mpc}^{-1}$}} 
\def\Kkms {\hbox{${\rm K\,km\,s}^{-1}$}}
\def\percc {$\hbox{{\rm cm}}^{-3}$}    %cm-3
\def\cmsq  {$\hbox{{\rm cm}}^{-2}$}    %cm-2
\def\cmsix  {$\hbox{{\rm cm}}^{-6}$}  %cm-6
\def\arcsec {\hbox{$^{\prime\prime}$}}
\def\arcmin {\hbox{$^{\prime}$}}
\def\ffam {\hbox{$\,.\!\!^{\prime}$}}
\def\ffas {\hbox{$\,.\!\!^{\prime\prime}$}}
\def\ffM {\hbox{$\,.\!\!\!^{\rm M}$}}
\def\ffm {\hbox{$\,.\!\!\!^{\rm m}$}}
\def\ffs {\hbox{$.\,\!\!^{\rm s}$}}
\def\HI  {\hbox{\ion{H}{i}}}
\def\HII {\hbox{HII}}
%
%   Greek and abbreviations for radio recomb lines etc
%
\def \AL {$\alpha $}    % gr. alpha
\def \BE {$\beta $}     % gr. beta
\def \GA {$\gamma $}    % gr. gamma
\def \DE {$\delta $}    % gr. delta
\def \EP {$\epsilon $}  % gr. epsilon
\def \alde {($\Delta \alpha ,\Delta \delta $)}
\def \MU {$\mu $}       % gr. mue
\def \TAU {$\tau $}     % gr. tau
\def \tapp {$\tau _{\rm app}$}
\def \tuns {$\tau _{\rm uns}$}
\def \RH {\hbox{$R_{\rm H}$}}         % OH main line ratio
\def \RT {\hbox{$R_{\rm \tau}$}}      % OH main tau  ratio
\def \BN  {\hbox{$b_{\rm n}$}}        % bn
\def \BETAN {\hbox{$\beta _n$}}       % beta factor
\def \TE {\hbox{$T_{\rm e}$}}         % Electron Temp.
\def \NE {\hbox{$N_{\rm e}$}}         % Electron Dens.
% molecules
%
\def\MOLH {\hbox{${\rm H}_2$}}                    %H2
\def\HDO {\hbox{${\rm HDO}$}}                     %HDO
\def\AMM {\hbox{${\rm NH}_{3}$}}                  %NH3
\def\NHTWD {\hbox{${\rm NH}_2{\rm D}$}}           %NH2D
\def\CTWH {\hbox{${\rm C_{2}H}$}}                 %C2H
\def\TCO {\hbox{${\rm ^{12}CO}$}}                 %12CO
\def\CEIO {\hbox{${\rm C}^{18}{\rm O}$}}          %C18O
\def\CSEO {\hbox{${\rm C}^{17}{\rm O}$}}          %C17O
\def\CTHFOS {\hbox{${\rm C}^{34}{\rm S}$}}        %C34S
\def\THCO {\hbox{$^{13}{\rm CO}$}}                %13CO
\def\WAT {\hbox{${\rm H}_2{\rm O}$}}              %H2O
\def\WATEI {\hbox{${\rm H}_2^{18}{\rm O}$}}       %H218O
\def\CYAC {\hbox{${\rm HC}_3{\rm N}$}}            %HC3N
\def\CYACFI {\hbox{${\rm HC}_5{\rm N}$}}          %HC5N
\def\CYACSE {\hbox{${\rm HC}_7{\rm N}$}}          %HC7N
\def\CYACNI {\hbox{${\rm HC}_9{\rm N}$}}          %HC9N
\def\METH {\hbox{${\rm CH}_3{\rm OH}$}}           %CH3OH
\def\MECN {\hbox{${\rm CH}_3{\rm CN}$}}           %CH3CN
\def\CH3C2H {\hbox{${\rm CH}_3{\rm C}_2{\rm H}$}} %CH3C2H
\def\FORM {\hbox{${\rm H}_2{\rm CO}$}}            %H2CO
\def\MEFORM {\hbox{${\rm HCOOCH}_3$}}             %HCOOCH3
\def\THFO {\hbox{${\rm H}_2{\rm CS}$}}            %H2CS
\def\ETHAL {\hbox{${\rm C}_2{\rm H}_5{\rm OH}$}}  %C2H5OH
\def\CHTHOD {\hbox{${\rm CH}_3{\rm OD}$}}         %CH3OD
\def\CHTDOH {\hbox{${\rm CH}_2{\rm DOH}$}}        %CH2DOH
\def\CYCP {\hbox{${\rm C}_3{\rm H}_2$}}           %C3H2
\def\CTHHD {\hbox{${\rm C}_3{\rm HD}$}}           %C3HD
\def\HTCN {\hbox{${\rm H^{13}CN}$}}               %H13CN
\def\HNTC {\hbox{${\rm HN^{13}C}$}}               %HN13C
\def\HCOP {\hbox{${\rm HCO}^+$}}                  %HCO+
\def\HTCOP {\hbox{${\rm H^{13}CO}^{+}$}}          %H13CO+
\def\NNHP {\hbox{${\rm N}_2{\rm H}^+$}}           %N2H+
\def\CHTHP {\hbox{${\rm CH}_3^+$}}                %CH3+
\def\CHP {\hbox{${\rm CH}^{+}$}}                  %CH+
\def\ETHCN {\hbox{${\rm C}_2{\rm H}_5{\rm CN}$}}  %C2H5CN
\def\DCOP {\hbox{${\rm DCO}^+$}}                  %DCO+
\def\HTHP {\hbox{${\rm H}_{3}^{+}$}}              %H3+ 
\def\HTWDP {\hbox{${\rm H}_{2}{\rm D}^{+}$}}      %H2D+
\def\CHTWDP {\hbox{${\rm CH}_{2}{\rm D}^{+}$}}    %CH2D+
\def\CNCHPL {\hbox{${\rm CNCH}^{+}$}}             %CNCH+
\def\CNCNPL {\hbox{${\rm CNCN}^{+}$}}             %CNCN+
%
% Abbreviations for T. Wiklind article
%
\def\In {\hbox{$I^{n}(x_{\rm k},y_{\rm k},u_{\rm l}$})}
\def\Iobs {\hbox{$I_{\rm obs}(x_{\rm k},y_{\rm k},u_{\rm l})$}}
\def\Ingl {I^{n}(x_{\rm k},y_{\rm k},u_{\rm l})}
\def\Iobsgl {I_{\rm obs}(x_{\rm k},y_{\rm k},u_{\rm l})}
\def\Pbgl {P_{\rm b}(x_{\rm k},y_{\rm k}|\zeta _{\rm i},\eta _{\rm j})}
\def\Pbgm {P(x_{\rm k},y_{\rm k}|r_{\rm i},u_{\rm l})}
\def\Pbgn {P(x,y|r,u)}
\def\Pugm {P_{\rm u}(u_{\rm l}|w_{\rm ij})}
\def\Pdem {P_{\rm b}(x,y|\zeta (r,\theta ),\eta (r,\theta ))} 
\def\Pden {P_{\rm u}(u,w(r,\theta ))}
\def\greekgl {(\zeta _{\rm i},\eta _{\rm j},u_{\rm l})}
\def\greekg1 {(\zeta _{\rm i},\eta _{\rm j})}

\def\ffam {\hbox{$\,.\!\!^{\prime}$}}
\def\ffas {\hbox{$\,.\!\!^{\prime\prime}$}}
\def\ffM  {\hbox{$\,.\!\!^{\rm M}$}}
\def\ffm  {\hbox{$\,.\!\!^{\rm m}$}}

\def\JB   {\,Jy\,beam$^{-1}$}                 %  \JB  : Jy beam^-1
\def\KB   {\,K\,beam$^{-1}$}                  %  \JB  : K beam^-1
\def\MJB  {\,mJy\,beam$^{-1}$}                %  \MJB : mJy beam^-1
\def\JBKS {\,Jy\,km\,s$^{-1}$\,beam$^{-1}$}   % \JBKS : Jy beam^-1 km s^-1
\newcommand{\Rvmax}{$R_{\rm {\mbox{v$_{\rm max}$}}}$}
\newcommand{\rvmax}{R_{\rm {\mbox v_{\rm max}}}}
\newcommand{\Rmax}{$R_{\rm max}$}
\newcommand{\Vmax}{$V_{\rm max}$}
\newcommand{\vmax}{V_{\rm max}}
\newcommand{\Vsys}{$V_{\rm sys}$}        %   \Vsys       :    V_sys
\newcommand{\Vhel}{$V_{\rm hel}$}
\newcommand{\Vlsr}{$V_{\rm lsr}$}
\newcommand{\vrot}{$v_{\rm rot}$}        %   \vrot       :    v_rot
\newcommand{\x}{\,$\times$\,}            %   \x          :    x
\newcommand{\HTWO}{H$_{\rm 2}$}          %   \HTWO      :     H_2
\newcommand{\HA}{H$_{\rm \alpha}$}       %   \HA        :     H_alpha
\newcommand{\MHTWO}{$M_{\rm H_{\rm 2}}$} %   \MHTWO      :    M_H2
\newcommand{\Msol}{M$_{\odot}$}          %   \Msol       :    solar mass unit
\newcommand{\Lsol}{L$_{\odot}$}          %   \Lsol       :    sol. lumin. unit
\newcommand{\LIR}{$L_{\rm IR}$}          %   \LIR        :    IR luminosity
\newcommand{\Tsys}{$T_{\rm sys}$}        %   \Tsys       :    Tsys
\newcommand{\TAS}{$T^*_{\rm A}$}         %   \TAS        :    T_a^stern
\newcommand{\COONE}{$^{12}$CO(1$-$0)}    %   \COONE      :   12^CO(1-0)
\newcommand{\COTWO}{$^{12}$CO(2$-$1)}    %   \COTWO      :   12^CO(2-1)
\newcommand{\COTRI}{$^{12}$CO(3$-$2)}    %   \COTRI      :   12^CO(3-2)
\newcommand{\LONE}{(1$-$0)}              %   \LONE       :    (1-0)
\newcommand{\LTWO}{(2$-$1)}              %   \LTWO       :    (2-1)
\newcommand{\LTRI}{(3$-$2)}              %   \LTRI       :    (3-2)
%
%\thesaurus{ 03                         % Extragalactic astronomy
%           (11.01.2                    % Galaxies: active
%            11.09.1 NGC\,3079          % Galaxies: individual: NGC\3079,
%            11.09.4                    % Galaxies: ISM
%            11.14.1                    % Galaxies: nuclei
%            13.19.1                    % Radio lines: galaxies
%       }

\title{ Red-shifted H$_2$O emission in NGC\,3079: More evidence for a pc-scale
        circumnuclear torus?
         \thanks{Based on observations with the 100-m telescope of the 
          MPIfR (Max-Planck-Institut f{\"u}r Radioastronomie) at Effelsberg}
       }

\author{Y. Hagiwara       \inst{1,2}
   \and C. Henkel         \inst{1}
   \and W.A. Sherwood     \inst{1}
   \and W.A. Baan         \inst{2}
       }

\institute{Max-Planck-Institut f{\"u}r Radioastronomie,
              Auf dem H{\"u}gel 69, D-53121 Bonn, Germany
   \and
           Westerbork Observatory, P.O. Box 2, NL-7990 AA Dwingeloo, 
              Netherlands
             }

\offprints{Y. Hagiwara, Dwingeloo}

\date{Received ... ; accepted ... }

\titlerunning{H$_2$O in NGC\,3079}

\authorrunning{Y. Hagiwara, C. Henkel, W.A. Sherwood and W.A. Baan }

%\markboth{Y. Hagiwara, C. Henkel, W.A. Sherwood, and W.A. Baan:
%             H$_2$O in NGC\,3079}
%         {Y. Hagiwara, C. Henkel, and W.A. Sherwood:
%             H$_2$O in NGC\,3079}

\abstract{
Using the Effelsberg 100-m telescope, sensitive measurements of the H$_2$O 
megamaser in NGC\,3079 are presented. During 2000 -- 2001, `high velocity' 
features are seen that are red-shifted up to 225\,\kms\ with respect to the 
systemic velocity of the galaxy ($V_{\rm LSR}$ $\sim$ 1120\,\kms). Symmetrically
bracketing the systemic velocity, the H$_2$O emission covers a velocity range 
of $\sim$450\,\kms\ with only one potential narrow gap ($\sim$20\,\kms) near 
the systemic velocity itself. Velocity drifts of individual components are not 
convincingly detected. It is shown that the presence of red-shifted emission 
and the absence of detectable velocity drifts are not inconsistent with the 
existence of a rotating circumnuclear maser disk at the very center of the galaxy. 
Significant differences in the overall line profile compared to NGC\,4258 
and a complex morphology of the radio continuum leave, however, space for scepticism.
         }

\maketitle

%--------------------------------------------------------------------------

\section{Introduction}

Recent single-dish and VLBI observations of luminous H$_2$O megamasers have been 
motivated by the discovery of an edge-on Keplerian sub-pc scale maser disk 
enshrouding a compact supermassive object at the nucleus of the LINER galaxy 
NGC\,4258 (e.g. Nakai et al. 1993, 1995; Haschick et al. 1994; Greenhill et al. 
1995; Miyoshi et al. 1995; Herrnstein et al. 1999). Circumnuclear disk structures 
traced by H$_2$O are also seen towards NGC\,1068, NGC\,4945, and the Circinus galaxy 
(e.g. Gallimore et al. 2001; Greenhill et al. 1997, 2000). H$_2$O megamasers therefore 
provide a unique probe to study the kinematics and the dynamical structure of the 
innermost regions of active galactic nuclei.

For the LINER or Seyfert 2 galaxy NGC\,3079 (see Sawada-Satoh et al. 2000) that 
contains one of the most luminous H$_2$O megamasers known to date (e.g. Henkel 
et al. 1984; Haschick \& Baan 1985; Haschick et al. 1990), arguments in favor of a 
circumnuclear disk have been less compelling. While the known maser components 
arise in the inner few parsecs of the galaxy, they do not align at right angles to the 
radio jet(s), almost all the detected emission is blue-shifted relative to the systemic 
velocity, and any velocity drift of individual H$_2$O features must be small 
($\la$4.0\,\kms\,yr$^{-1}$; Nakai et al. 1995; Baan \& Haschick 1996; Trotter et al. 1998; 
Satoh et al. 1999; Sawada-Satoh et al. 2000). 

In this letter we therefore report single-dish observations of the 22\,GHz water vapor 
line from NGC\,3079, (1) to look for the missing red-shifted velocity components and (2) to 
search for velocity drifts in the near systemic features that would indicate centripetal 
acceleration in a rotating circumnuclear disk.

\section{Observations}

Observations of the $J_{\rm K_a K_c} = 6_{16}-5_{23}$ H$_2$O maser line (rest frequency: 
22.23508\,GHz) were made with the MPIfR 100-m radio telescope at Effelsberg between 
February 1994 and December 2001. Until 1998, a single channel K-band maser receiver was 
employed in a position switching mode with a system temperature of $T_{\rm sys}$ $\sim$ 
225\,K on a main beam brightness temperature ($T_{\rm mb}$) scale. An autocorrelator 
provided bandwidths of 50, 25 or 12.5\,MHz with 1024 spectral channels, yielding channel 
spacings of 0.66, 0.33 or 0.16\,\kms. Since 2000, we used a dual channel K-band HEMT 
receiver in a dual beam switching mode with a beam throw of 2$'$ and a switching 
frequency of 1\,Hz. After averaging the two orthogonally polarized signals, $T_{\rm sys}$ 
$\sim$ 180\,K. Four autocorrelator backends were employed for each receiver channel. 
Each spectrum had a bandwidth of 40\,MHz and 512 channels, yielding a channel spacing of 
1.05\,\kms. 

\begin{table}[t]
\caption[Epochs]{Effelsberg 100-m observations log$^{\rm a)}$}
\begin{center}
\begin{tabular}{llr|llr}
Epoch&     &        & Epoch&     &      \\
     &     &        &      &     &      \\
1994 & Feb &      9 & 2000 & Mar & 18   \\
1995 & Sep & 16--17 & 2000 & Oct & 14   \\
1996 & Sep & 21--23 & 2000 & Dec & 21   \\
1997 & Apr &  5--6  & 2000 & Mar & 13   \\
1997 & Nov &      5 & 2001 & Apr & 22   \\
1998 & Feb &      1 & 2001 & May &  9   \\
1998 & May &      8 & 2001 & Dec &  5   \\
1998 & Jun &     27 &      &     &      \\
1998 & Aug &      1 &      &     &      \\
\end{tabular}
\end{center}
{\footnotesize 
a) Several tunings and bandwidths were employed during most epochs.
As a consequence, noise levels and channel spacings are not 
uniform across the observed velocity ranges. 1$\sigma$ noise 
levels vary from 200\,mJy (channel spacing 1.32\,\kms) in September 
1996 to 7\,mJy (channel spacing 1.05\,\kms) in December 2001. Velocity 
resolution $\sim$ 1.25\,$\times$\,channel spacing. For direct
access to 64 flux calibrated spectra, contact chenkel@mpifr-bonn.mpg.de.}
\end{table}

Amplitude calibration was based on measurements of the 22\,GHz 
continuum flux of 3C\,286 (see Baars et al. 1977; Ott et al. 1994). Pointing 
measurements toward nearby sources (in most cases DA\,251) were made once per hour. 
The resultant pointing accuracy was $<$ 8'', that should be compared with 
the full width to half power beam size of 40$''$. Calibration uncertainties 
are estimated to be $\pm$15\%. 

\section{Results}

H$_2$O spectra were taken during 16 epochs (Table 1; no changes in the maser 
profiles were seen {\it within} any of these observing  periods). Several 
H$_2$O features could be traced over the entire monitoring period. The 
most prominent component, detected at $\sim$956\,\kms, showed a peak 
flux density of 2.5--4.5\,Jy. While the peak velocity is observed to 
drift from about 955.0 to 956.8\,\kms\ (estimated error of individual 
measurements: 0.3\,\kms), this drift is not systematic and appears to be 
caused by a multitude of individual, variable subcomponents. During times 
with lower peak velocity, the profile tends to show a `shoulder' at the 
high velocity wing of the line (e.g., on Nov. 5, 1997 and Feb. 1, 1998) 
and vice versa (Dec. 21, 2000, and Apr. 22, 2001). Notable were two flares 
of more red-shifted components that exceeded 2\,Jy: The 979\,\kms\ 
component with 0.7\,Jy during Feb. 1994 to Apr. 1997 (see Nakai et al. 1995 
for earlier spectra) reached 0.9\,Jy in Nov. 1997, 1.5\,Jy in Feb. 1998, and 
2.2\,Jy in May 1998 (Fig.\,1a); it then decreased to 1.8\,Jy in June 1998, 
1.25\,Jy in Aug. 1998, and 0.35\,Jy since Oct. 2000. The 1017\,\kms\ velocity 
component, observed between Sep. 1996 and Nov. 1997 at 0.1--0.5\,Jy, reached 
a peak flux density of 2.5\,Jy in Mar. 2000 (Fig.\,1b); flux densities decreased 
to 1.0\,Jy in Oct. 2000 and to 0.1\,Jy in May 2001. Emission near the systemic 
velocity of the galaxy (1080\,\kms\ $<$ $V_{\rm LSR}$ $<$ 1160\,\kms) 
remained faint and not a single narrow velocity component could be traced 
during several consecutive observing epochs. 

\begin{figure}
\vspace{-2.2cm}
\hspace{0.0cm}
\psfig{figure=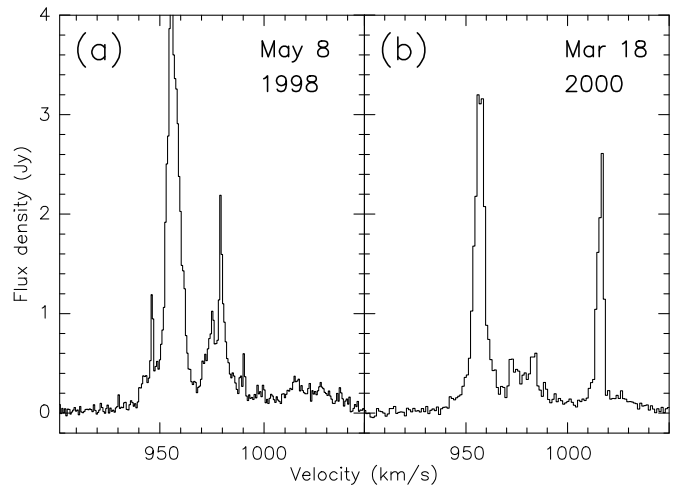,width=13.9cm,angle=-90.}
\vspace{-1.3cm}
\caption[fig1]{The (a) 979 and (b) 1017\,\kms\ features flaring beyond the 2\,Jy level
(channel spacings: 0.66 and 1.05\,\kms). Velocity scale throughout the paper: Local 
Standard of Rest (LSR), following the radio astronomical definition of velocity 
(see e.g. Trotter et al. 1998). $V_{\rm LSR}$ = $V_{\rm HEL}$ + 3.23\,\kms.}
\label{fig1}
\end{figure}

Fig.\,2 shows a spectrum of near systemic ($V_{\rm sys}$$\sim$1120\,\kms) and 
red-shifted ($V$$>$$V_{\rm sys}$) features that were detected since March 2000. 
More than 10 distinct narrow components appear at velocities in excess of 
1100\,\kms\ (Table 2). We conclude that the H$_2$O maser emission extends over 
a radial velocity range of $\sim$450\,\kms. The integrated luminosity of the 
previously known blue-shifted components remains approximately constant. This 
may also hold for the systemic and red-shifted features shown in Fig.\,2. A 
detection of these prior to 2000 would have been difficult in view of technical 
improvements that occurred at Effelsberg in 1999. Sensitivity also limits the 
studies by Nakai et al. (1995; we do not confirm the weak 764 and 791\,\kms\ 
components) and Baan and Haschick (1996). However, the 1123 and 1190\,\kms\ 
components observed by Trotter et al. (1998) in Jan. 1995, the flaring 1192\,\kms\ 
feature ($\sim$0.4\,Jy) detected by Nakai et al. (1995) in Apr.-May 1995, 
and a 1201\,\kms\ feature detected by us in Sept. 1995 (Fig.\,3) indicate 
the presence of isolated red-shifted maser features as early as half a dozen 
years ago. 

\begin{figure}
\vspace{-3.2cm}
\hspace{+0.0cm}
\psfig{figure=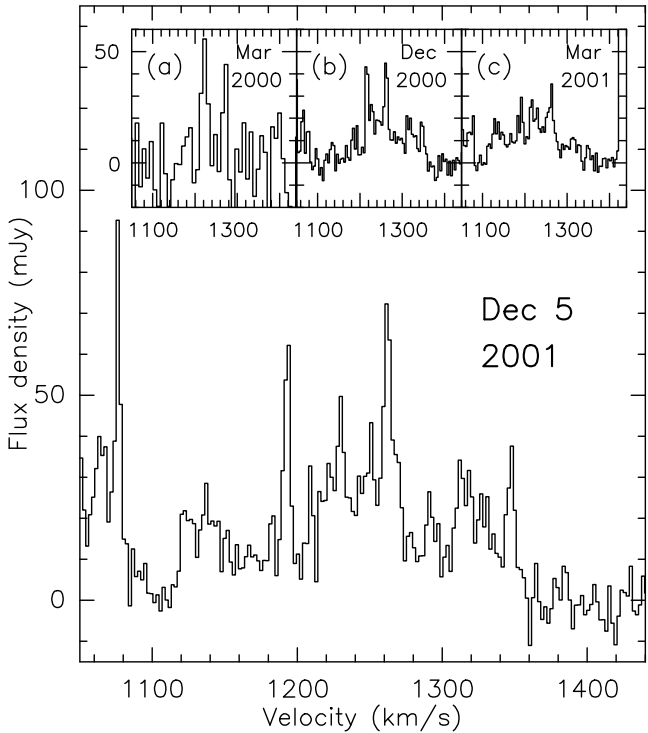,width=18.7cm,angle=-90.}
\caption[fig1]{Near systemic and red-shifted H$_2$O maser features toward NGC\,3079. 
Channel spacing: 2.1\,\kms\ (main spectrum), 4.2\,\kms\ (inserts (b) and (c)), and
8.4\,\kms\ (insert (a)). The inserted spectra (insert (c) includes a few bad channels 
at the edge of the band) show an initially dominant fading 1222\,\kms\ component, 
while the flux of the 1192\,\kms\ component increases. The brightest components have 
an isotropic luminosity of $\sim$0.5\,L$_{\odot}$. Since the zero level is determined 
by a linear baseline derived from velocities $V$ $<$ 920 and $>$ 1360\,\kms, the absence 
of H$_2$O emission near 1100\,\kms\ is not entirely certain.} 
\label{fig2}
\end{figure}

\section{Discussion}

In NGC\,4258, Nakai et al. (1993) detected red- and blue-shifted H$_2$O `satellites'
symmetrically displaced by approximately $\pm$900\,\kms\ from the systemic velocity. 
This suggested the presence of a circumnuclear disk that was later confirmed (see 
Sect.\,1). Is the discovery of a multitude of near systemic and red-shifted features 
in NGC\,3079 also hinting at a circumnuclear disk? Emission is seen at least between 
928 and 1352\,\kms\ (Fig.\,3). Assuming the presence of a symmetric edge-on circumnuclear 
disk, the systemic velocity of the nuclear region then is $\sim$1140\,\kms. Including 
the 896\,\kms\ feature apparent in the early spectra of Henkel et al. (1984) and 
Haschick \& Baan (1985) and tentatively also seen in Dec. 2000 and Dec. 2001, 
$V_{\rm sys}$ $\sim$ 1124\,\kms. Is this consistent with the systemic velocity of the 
galaxy?

From the nuclear molecular gas traced by the CO $J$ = 1--0 line, Irwin \& Sofue (1992)
deduced a systemic velocity of $V_{\rm sys}$ = 1050$\pm$10\,\kms. From an \hbox{\ion{H}{i}} 
line profile, Irwin \& Seaquist (1991) obtained a midpoint at the 20\% peak level of 
1123$\pm$10\,\kms\ and 1126\,\kms\ from a fit to the moment map. Three dimensional 
modeling of their \hbox{\ion{H}{i}} data cube yields $V_{\rm sys}$ = 1116\,\kms. The 
rough agreement of CO, \hbox{\ion{H}{i}}, and H$_2$O velocity centroids implies a 
nuclear systemic velocity of $\sim$1120\,\kms\ and suggests that we have seen, for 
the first time, all the stronger 22\,GHz H$_2$O maser components in NGC\,3079. A 
nuclear systemic velocity of 1230\,\kms\ as suggested by Satoh et al. (1999) is not 
supported by our H$_2$O data.

\begin{table}[t]
\caption[Velocities]{Velocities of distinct near systemic or red-shifted features 
observed on December 5, 2001 (see Fig.\,2)}
\begin{center}
\begin{tabular}{c|c}
   $V$(\kms)   &    $V$(\kms)    \\
               &                 \\
1062.7$\pm$0.4 & 1209.3$\pm$ 0.3 \\
1066.8$\pm$1.4 & 1229.8$\pm$ 0.4 \\
1076.5$\pm$0.2 & 1251.1$\pm$ 0.1 \\
1123.0$\pm$0.9 & 1262.1$\pm$ 0.2 \\
1137.7$\pm$1.4 & 1291.5$\pm$ 0.9 \\
1182.6$\pm$0.4 & 1314.4$\pm$ 1.0 \\
1192.9$\pm$0.5 & 1347.6$\pm$ 0.4 \\
\end{tabular}
\end{center}
\end{table}

Nevertheless, the maser emission is far from perfectly symmetric: (1) All bright components 
($>$0.5Jy) are blue-shifted with respect to the systemic velocity which is inconsistent with 
the situation in NGC\,4258 and NGC\,1068 as well as with the spiral shock model proposed 
by Maoz \& McKee (1998). (2) The systemic velocity obtained from model fits (Irwin \& 
Seaquist 1991) lies at the edge of that narrow velocity interval ($\Delta V$$\sim$20\,\kms, 
see Fig.\,2) that appears to be devoid of H$_2$O emission (the only such interval over the 
entire H$_2$O velocity range observed by us). Interpreted in terms of the paradigm established 
for NGC\,4258, this may hint at a lack of nuclear 22\,GHz radio continuum emission at 
the very center of the putative masering torus. (3) In a well ordered edge-on circumnuclear 
disk, small line-of-sight velocity gradients that stimulate maser emission should only be present 
near the tangential points and near the front and back side of the disk. A continuous coverage 
of the velocity range as observed in NGC\,3079 is not expected. (4) None of the four systemic 
velocities quoted above provides a particularly high number of red- and blue-shifted components 
that match each other with respect to velocity ($|V_{\rm red}-V_{\rm sys}|$ $\sim$ 
$|V_{\rm blue}-V_{\rm sys}|$). Note that the following discussion, does not sensitively depend 
on the exact choice of $V_{\rm sys}$ as long as it is in the range 1100--1150\,\kms.

Most blue-shifted 22\,GHz H$_2$O velocity components of NGC\,3079 are already known 
to show no significant velocity drift (e.g. Fig.\,7 of Nakai et al. 1995; 
Fig.\,6 of Baan \& Haschick 1996). The slow drift of $\sim$0.4\,\kms\,yr$^{-1}$ suggested
by Baan \& Haschick (1996) for the 945, 951, and 1015\,\kms\ components ($V$ refers 
to epoch 1994.0) is not expected if a circumnuclear torus is present (e.g. Miyoshi et 
al. 1995). Our data for the 945\,\kms\ component (945.7 and 946.2\,\kms\ in Feb. 1994 
and 1998, and 946.6 and 946.5\,\kms\ in May and June 1998; errors derived from Gaussian 
fits are $\sim$0.3\,kms) are not contradicting Baan \& Haschick (1996), although a group 
of features with varying amplitudes could also simulate the drift. The 1017\,\kms\ 
component (Fig.\,1b) fluctuates between 1014 and 1018\,\kms\ and does not support a regular
drift, although it is likely related to the 1015\,\kms\ component of Baan and Haschick (1995). 
The 951\,\kms\ component was not seen by us. Satoh et al. (1999) and Sawada-Satoh et al. 
(2000) report a drift of 3.7$\pm$0.6\,\kms\,yr$^{-1}$ for a maser feature near 1190\,\kms. 
We find a group of features (total width: $V_{\rm tot}$$\sim$5--6\,\kms) centered at 
1192.9\,\kms, both on Mar 3. and Dec. 5, 2001. Thus the drift appears to be well below 
our estimated detection level of 1.5\,\kms\,yr$^{-1}$ for a time interval of 9 months. 
In view of the long lifetime inferred for some of the blue-shifted maser lines (Sect.\,3 
and Fig.\,3), the 1192\,\kms\ feature reported by Nakai et al. (1995) is likely part of 
this component, reducing the velocity drift to well below 1.0\,\kms\,yr$^{-1}$.

\begin{figure}
\vspace{-0.0cm}
\hspace{+0.0cm}
\psfig{figure=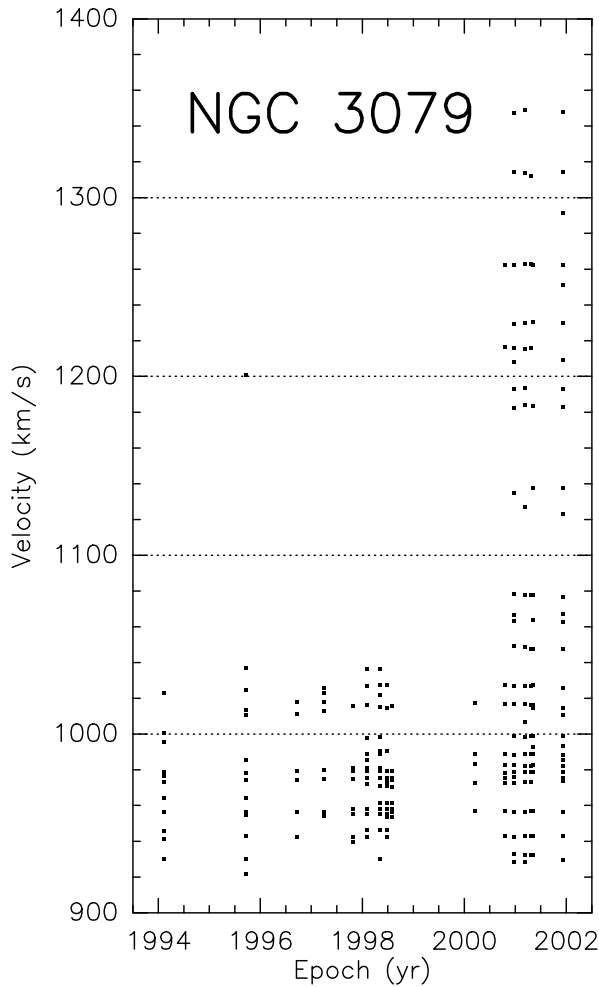,height=13.5cm,angle=-90.}
\caption[fig1]{Individual velocity components of the 22\,GHz H$_2$O spectrum 
obtained with an accuracy $\la$1\,\kms\ for the blue-shifted components
and $\sim$1\,\kms\ for the redshifted ones.} 
\label{fig3}
\end{figure}

The overall lineshape of the H$_2$O profile is quite distinct from that of NGC\,4258
and the morphology of the radio continuum (T. Krichbaum, priv. comm.) is more complex. In 
spite of the presence of a roughly linear ridge of blue-shifted H$_2$O masers, the existence 
of an associated nuclear disk is thus not certain (see e.g. Trotter et al. 1998 for alternative
scenarios). Sawada-Satoh et al. (2000) proposed a thick maser disk with continuum source `B' 
at the nucleus and the 1190\,\kms\ feature on its near side (their Fig.\,6). Since we see no 
velocity drift near 1190\,\kms\ and since this is not the systemic velocity, however, there 
is no need to put this component at the near side of the putative disk. We therefore tend 
to favor the disk scenario outlined by Trotter et al. (1998) with the nucleus of the galaxy 
being located between continuum components A and B (their Fig.\,7). In this latter scenario, 
expected velocity drifts would be consistent with our upper limits. The nucleus would contain 
a few 10$^{6}$\,M$_{\odot}$ within $R$ $\sim$ 10\,mas ($\sim$0.7\,pc; assumed rotation velocity 
$V_{\rm rot}$$\sim$100--225\,\kms), the centripetal acceleration and proper motion of the 
near-systemic components would be difficult to detect ($<$0.1\,\kms\,yr$^{-1}$ and 
$\sim$1\,$\mu$as\,yr$^{-1}$), and most of the red-shifted masers would be located 10--15 mas 
south of the blue-shifted ones (some masers may be associated with the jet (e.g. Trotter et al. 
1998)). Sensitive VLBI data are needed to examine this picture and to obtain definite information 
on the relative location of blue- and red-shifted maser features and continuum sources in the 
very center of NGC\,3079.

\begin{acknowledgement}

We wish to thank M. Inoue and L.J. Greenhill for critically reading the manuscript.

\end{acknowledgement}

{}

\end{document}